# Enhancing Covid-19 Decision-Making by Creating an Assurance Case for Simulation Models

Ibrahim Habli
Department of Computer Science
University of York
York, UK

Rob Alexander
Department of Computer Science
University of York
York, UK

Richard Hawkins
Department of Computer Science
University of York
York, UK

Mark Sujan
Warwick Medical School
University of Warwick
Coventry, UK

John McDermid
Department of Computer Science
University of York
York, UK

Chiara Picardi
Department of Computer Science
University of York
York, UK

Tom Lawton
Department of Anaesthesia and Critical Care
Bradford Teaching Hospitals
NHS Foundation Trust
Bradford, UK

**Abstract.** Simulation models have been informing the COVID-19 policy-making process. These models, therefore, have significant influence on risk of societal harms. But how clearly are the underlying modelling assumptions and limitations communicated so that decision-makers can readily understand them? When making claims about risk in safety-critical systems, it is common practice to produce an assurance case, which is a structured argument supported by evidence with the aim to assess how confident we should be in our risk-based decisions. We argue that any COVID-19 simulation model that is used to guide critical policy decisions would benefit from being supported with such a case to explain how, and to what extent, the evidence from the simulation can be relied on to substantiate policy conclusions. This would enable a critical review of the implicit assumptions and inherent uncertainty in modelling, and would give the overall decision-making process greater transparency and accountability.

## INTRODUCTION

The epidemiological simulation models that have been informing the COVID-19 policy-making process should be viewed as safety-critical systems. These models have direct and significant influence on the policies and decisions that aim to reduce the risk posed by the virus to public health [1]. However, a recent systematic review of 31 diagnostic and prediction models for COVID-19 concluded that, at present, none of these models could be recommended for practical use to inform critical policy decisions [2]. It is, therefore, vital that decision-makers are aware of the assumptions made in these models and that they can reflect on the limitations of the models in relation to practical decisions about the management of the pandemic.

Similar to engineered safety-critical systems, e.g. flight control software or pacemakers, the rigour and transparency with which these simulation models are developed should be proportionate to their criticality to, and influence on, public health policy - this is true for COVID-19 but also holds for other models used to support such critical decision-making. In safety-critical systems engineering it is common practice to produce an assurance case — a structured, explicit argument supported by evidence [3]. Such cases are a primary means by which confidence in the safety of the system is communicated to, and scrutinised by, the diverse stakeholders, including regulators and policy makers. We believe it is important to support the COVID-19 simulation models with an assurance case that explains how, and the extent to which, the resulting evidence supports and substantiates the policy conclusions. We argue that such a case has the potential to enable a wider understanding, and a critical review, of the expected benefits, limitations and assumptions that underpin the development of the simulation models and the extent to which these issues, including the different sources of uncertainty, are considered in the policy decision-making process.

## ASSURANCE CASES IN SAFETY-CRITICAL SYSTEMS

The use of assurance cases is a long-established practice in the safety-critical domain. Particularly in the UK, the development of an assurance case is a mandatory requirement in key sectors such as defence, nuclear and rail [4]. More pertinently, in the NHS, compliance with the clinical safety standards DCB0129 and DCB0160 requires an assurance case for Health IT systems [5]. An assurance case may consider different critical properties of a system. In this paper, we focus on safety.

An assurance case is primarily used to communicate, support and critically evaluate a safety claim about risk-based decisions to commission a system or a service. Data from modelling, simulation, testing and in-service usage provides the evidence base for such a claim. However, this evidence is rarely conclusive. It entails different sources of uncertainty and hinges on technical, organisational and social assumptions. Further, in making risk-based decisions, tradeoffs are inevitable, e.g. between safety, privacy and costs. Justifications for these decisions have to be communicated to and accepted by the relevant stakeholders, e.g. by individuals or groups whose privacy might be reduced in return for an improvement in safety. To this end, a structured argument is used to explain the extent to which the evidence supports the safety claim, given the many sources of uncertainty, assumptions and tradeoffs. The argument is structured in the sense that it should make these issues and the way in which they relate to each other explicit for the different stakeholders to critically review, modify, accept or reject. The more complex and novel the system and its context are, the more significant the role of the assurance argument is in informing the risk-based decision-making process.

Healthcare and public health interventions form a complex and adaptive set of interacting systems in which risk-based and evidence-based decisions would benefit from explicit and clear explanation by means of structured arguments. A study by the Health Foundation on the use of assurance cases in healthcare highlighted some potential benefits [6], primarily:
1. promoting structured thinking about risk among clinicians and fostering multidisciplinary communication about safety;
2. integrating evidence sources;
3. aiding communication among stakeholders; and
4. making the implicit explicit.

# ASSURANCE CASES FOR EVIDENCE-BASED COVID-19 POLICY

As a highly salient example, take Report 9 by the Imperial College COVID-19 Response Team ("the impact of Non-Pharmaceutical Interventions (NPIs) on the reduction of COVID-19 mortality and healthcare demand") [7]. This is an example of microsimulation providing primary evidence which has significant policy implications. We can view a policy assurance case as an integration of the following, as illustrated in Figure 1:

A. Data from microsimulation providing *scientific evidence*;
B. *Scientific claims*, often referred to as "scientific advice", concerning the effect of the different public health strategies based on the evidence; and
C. *Policy claims* concerning the chosen public health strategy based on the scientific claims, but also taking into account national values, policy goals, etc.

The relationships between the above can be established through the following arguments:

D. *Scientific argument* explaining the extent to which the microsimulation evidence (A) supports the scientific claims (B);
E. *Policy argument* explaining the extent to which the scientific claims (B) are sufficient to support the policy claims (C); and
F. *Confidence argument* in the trustworthiness of the microsimulation evidence (A).

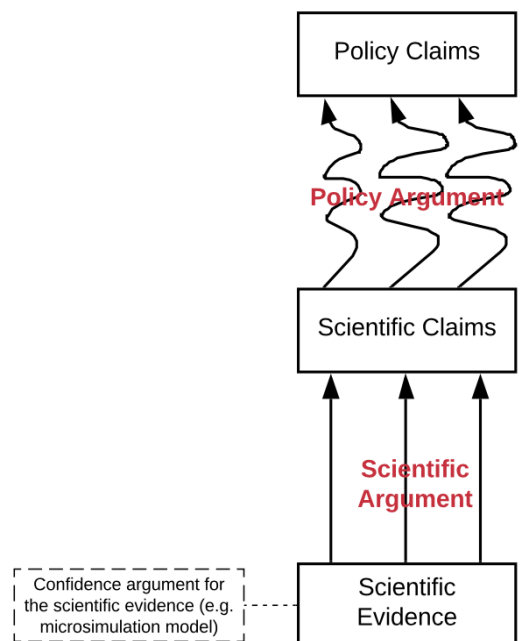

Figure 1: Overall Assurance Case Structure for a COVID-19 Policy

**Scientific Evidence based on Microsimulation**

Simulation models are engineering artefacts. As such, they should be systematically specified, implemented and tested. The rigour with which this is performed should be proportionate to the criticality of these models to the decision making process. For example, the COVID-19 model used by the Imperial team is based on a modified individual-based simulation that was developed to support pandemic influenza planning. Models can be for a specific purpose and therefore a confidence argument would need to justify the suitability of the model for the new context, including the continued validity of the original parameters. This is important since ad-hoc reuse and modification have been associated with catastrophic accidents in other safety-critical domains (e.g. the recent Boeing 737 Max accidents [8]). As Thimbleby recently argued [9], the quality of the software design and code of the simulator is an important factor, particularly its amenability to inspection and testing. For instance, Neil Ferguson, the lead author of the Imperial report, stated the following: "*For me the code is not a mess, but it's all in my head, completely undocumented. Nobody would be able to use it... and I don't have the*

*bandwidth to support individual users''* [10]. In a safety-critical context, this would significantly undermine confidence in the simulation results. We can note that this is defensible, in context — the Imperial team was working under tight timescales and (longer-term) it does plan to make the simulation program publicly available. We hope that this will be combined with the actual source code to enable wider replication and evaluation of the evidence.

The validity of the simulation results hinges on large uncertainties and many societal assumptions, e.g. about population behavioural changes. In large part, this is because COVID-19 is a novel virus, with relatively little reliable data on transmission rates. The developers of the model made many of these assumptions explicit through listing the corresponding parameters and where data exists to support the chosen parameter values. This should enable an independent assessment and evolution of the model. For example, the report states an assumption that 30% of COVID-19 patients who are hospitalised will require critical care (invasive mechanical ventilation) based on early reports from cases in the UK, China and Italy. We now know that this was a significant overestimate due to a combination of miscommunication ("critical care" in many other countries includes non-invasive measures such as continuous positive airway pressure devices) and the effects of the initial official UK advice to "intubate early".

**Scientific Claim and Argument**

Given the novelty of the virus and the large uncertainties around the design of the model and its underpinning data, the transition from the simulation results to the overall scientific claim, i.e. scientific advice or conclusions, is not straightforward [11]. We recreated the scientific argument using a structured argumentation notation, the Goal Structuring Notation (GSN) [12]. GSN is widely used in safety-critical domains for creating structured assurance arguments. Figure 2 shows a simplified example of how a structured argument can be used to capture part of the scientific argument through identifying the claims that are made, the evidence that supports those claims, and the relationships between them. Structured arguments also help to ensure that the key assumptions that are made are documented, e.g. exclusion of economic and non-COVID-19 health implications.

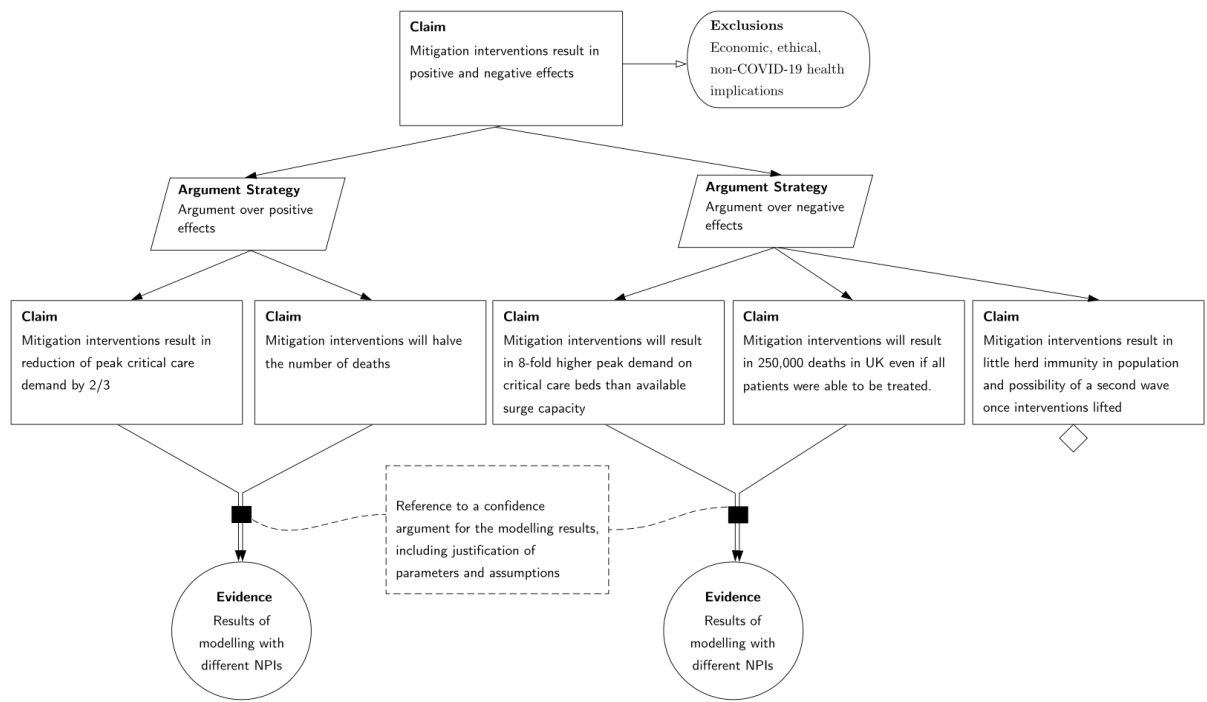

Figure 2: An example of part of a structured scientific argument for COVID-19 Simulation

In Figure 2, the results of modelling for different NPIs are used as evidence to support claims about the positive and negative effects of a COVID-19 mitigation strategy. A confidence argument in the trustworthiness §of the microsimulation evidence is developed further, referenced but not shown in Figure 2, and considers justification of the way in which the model was adapted and tested, including the choice of parameters.

**Policy Claim and Argument**
Moving from scientific advice and evidence to a policy decision requires that policy makers consider assumptions, risk acceptance beliefs and tradeoffs (such as between economic and medical impact) that are not often direct and amenable to rigorous scientific examination [11]. The transition from a scientific claim to a policy one should therefore involve a complex and diverse policy argument that builds on the scientific claims, but also brings to bear these additional considerations [13]. Imperial College Report 9 contains some explicit policy claims, but it does not contain a policy argument [7].

A good policy argument should justify the reliance on particular sources of scientific advice and models, and acknowledge the extent to which the underlying sources of uncertainty in the evidence were considered. Alternative scientific claims based on different (potentially conflicting) models should also be considered where available. The policy argument should make clear how tradeoffs were made and how evidence concerning the economic, legal and ethical implications of the chosen policy was generated and appraised. In the COVID-19 context, such evidence should also incorporate an estimation of non-COVID-19 health harms, e.g. potential delays in cancer diagnosis and treatment.

Highlighting the different aspects of the policy argument should ensure clarity about different accountabilities: (1) the accountability of the scientists to base their scientific advice on data and results that have been generated in a trustworthy manner; and (2) the accountability of the policy makers to appraise the different items of evidence and clarify the basis on which the policy was established.

## CONCLUSIONS AND RECOMMENDATIONS
Our society is currently placing great weight on simulation models of COVID-19 effects. Although such models are essential for dealing with the pandemic, it is hard to know which we should trust, to what extent, under what conditions. We need, therefore, to make an interdisciplinary effort to understand these models, and to support that effort we should use assurance cases to capture our arguments of validity.

In such an effort, epidemiologists and health data scientists will have a central role, but they will need support from software engineers, including those with safety-critical software experience. Working together, such collaborations will be able to create standards for the developing, testing and maintaining these models in a consistent, rigorous and auditable manner. They will be able to build assurance cases that communicate the uncertainty, assumptions and tradeoffs to a wide variety of stakeholders. This knowledge will then aid policymakers in using pandemic models in exactly the ways that they are useful, and not in the ways that they are not.